\documentclass[twocolumn]{revtex4}

\usepackage{amsmath}
\usepackage{graphicx}
\usepackage{color}
\usepackage[lofdepth,lotdepth,caption=false]{subfig}

\begin{document}

\title{Optimized pulse shapes for a resonator-induced phase gate}
\author{Andrew W. Cross}
\email{awcross@us.ibm.com}
\affiliation{IBM T. J. Watson Research Center, 1101 Kitchawan Road, Yorktown Heights, New York 10598}

\author{Jay M. Gambetta}
\email{jmgambet@us.ibm.com}
\affiliation{IBM T. J. Watson Research Center, 1101 Kitchawan Road, Yorktown Heights, New York 10598}

\date{11 November 2014}


\begin{abstract}
The resonator-induced phase gate is a multi-qubit controlled-phase gate for fixed-frequency superconducting qubits. Through off-resonant driving of a bus resonator, statically coupled qubits acquire a state-dependent phase. However, photon loss leads to dephasing during the gate, and any residual entanglement between the resonator and qubits after the gate leads to decoherence. Here we consider how to shape the drive pulse to minimize these unwanted effects. First, we review how the gate's entangling and dephasing rates depend on the system parameters and validate closed-form solutions against direct numerical solution of a master equation. Next, we propose spline pulse shapes that reduce residual qubit-bus entanglement, are robust to imprecise knowledge of the resonator shift, and can be shortened by using higher-degree polynomials. Finally, we present a procedure that optimizes over the subspace of pulses that leave the resonator unpopulated. This finds shaped drive pulses that further reduce the gate duration. Assuming realistic parameters,  we exhibit shaped pulses that have the potential to realize $\sim 212$ ns spline pulse gates and $\sim 120$ ns optimized gates with $\sim 6\times 10^{-4}$ average gate infidelity. These examples do not represent fundamental limits of the gate and, in principle, even shorter gates may be achievable.
\end{abstract}

\maketitle

\section{Introduction}

As the lifetime of superconducting qubits continues to grow and more complex multi-qubit control becomes feasible \cite{chow14,barends14,saira14,devoret13,corcoles14}, there is a continuing need to explore two-qubit gates with the potential to have sufficiently high fidelity to enable fault-tolerant quantum error-correction \cite{rh07,aliferis06}. In addition, gates require sufficiently low leakage error rates \cite{aliferis07,suchara14} and cross-talk \cite{gambettaaddr12,fowlercorr14}. Gates for fixed-frequency qubits must be robust to qubit frequency imprecision and spectral crowding. It is an ongoing challenge to identify a gate that satisfies all of these constraints and operates within a physically reasonable range of parameters.

Entangling gates for superconducting qubits fall into two broad classes based on whether or not the gates rely on dynamic flux tunability. Flux-tunable gates include the direct-resonant iSWAP (DRi) \cite{bialczak10,dewes12}, the dynamical c-Phase (DP) \cite{dicarlo09,strauch03,yamamoto10}, and other variants \cite{niskanen07,bialczak11}. These gates have the advantage of tuning the interaction on and off so that gates are fast when the interaction is on and cross talk is low when the interaction is off. On the other hand, flux tunability can introduce noise that reduces coherence time \cite{yoshihara06}, risk tuning the qubit through a resonance with a two-level system \cite{martinis05} (although see \cite{andersen14}), and require more complex circuitry and control lines. Microwave-only gates do not rely on flux tunability and allow qubit frequencies to remain fixed. This class of gates includes the resonator sideband induced iSWAP (RSi) \cite{leek09}, the cross-resonance (CR) gate \cite{corcoles13,paraoanu06,rigetti10,chowcr11}, the bSWAP gate \cite{poletto12}, and the microwave-activated c-Phase (MAP) gate \cite{chowmap13}. Qubits with fixed frequencies can be designed to improve coherence times, and two-qubit control becomes analogous to single qubit gate controls. Drawbacks include challenges to designing coupling strengths to achieve fast gates compatible with long coherence times and low cross-talk.

The resonator-induced phase gate is a microwave-only controlled-Phase gate where fixed-frequency transmons are statically coupled to the same driven bus resonator \cite{gambetta12,paik14}. The main limitation is that photons exist in the resonator during the gate, and large qubit-bus coupling strengths can lead to significant qubit dephasing \cite{gambetta06} unless the drive is appreciably detuned from the dressed resonator frequency. However, the gate tolerates potentially large variability in the qubit frequencies and is not expected to produce significant leakage errors, since the drive is tuned relatively far above the bus frequency. The approach of driving the bus also generalizes to multi-qubit systems if qubits share the same bus, although this may require more complex pulse shapes.

Our main goal here is to explore theoretical drive pulse shapes for the resonator-induced phase gate that minimize gate error while making physically reasonable assumptions. We derive relatively simple close-form solutions for the gate, including its entangling and dephasing rates, and validate these solutions against direct numerical evolution under a master equation. Using the close-form solutions, we propose spline pulse shapes that reduce residual qubit-bus entanglement, are robust to imprecise knowledge of the resonator shift, and can be shortened by using higher-degree polynomials. Dissipation motivates the goal of further reducing the gate duration, but short pulses populate the bus resonator. To overcome this problem, we present a new procedure that optimizes over the subspace of pulses that leave the resonator unpopulated. This procedure finds numerical shaped drive pulses that further reduce the gate duration.

The paper is organized as follows. Section~\ref{sec:rip} introduces the resonator-induced phase gate for two non-linear oscillators coupled by a bus resonator. Section~\ref{sec:constant} analyzes the rates of entanglement generation and measurement-induced dephasing when the bus resonator is driven by a tone of constant amplitude. Section~\ref{sec:spline} then describes a spline shaped pulse and shows that such a shape can ensure that the bus resonator returns approximately to its ground state after the gate. Finally, Section~\ref{sec:nullspace} presents a new optimization technique wherein solutions are restricted to the nullspace of a linear operator to ensure that the resonator returns precisely to its ground state. The optimized pulse shapes suggest a general form for fast high fidelity gates.

\section{Resonator-induced phase gate}\label{sec:rip}

\begin{figure}
\includegraphics[width=8cm]{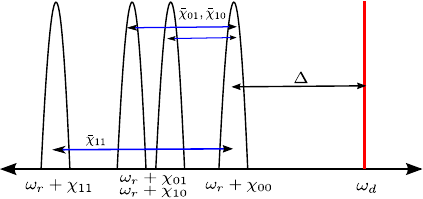}
\caption{
(color online) 
The bus resonance is shifted up in frequency by $\chi_{jk}$ when the qubits are placed in the state $|jk\rangle$. When the bus resonator is driven by a tone at $\omega_d$ that is detuned by $\Delta>0$ from the ground state bus resonance, the resonator responds differently for each qubit state. The response is a function of the detuning $\Delta$ and the shifts $\bar{\chi}_{jk}=\chi_{00}-\chi_{jk}$.\label{fig:peaks}}
\end{figure}

We consider two nonlinear oscillators coupled by a bus resonator that is driven by a single tone with a  shaped envelope. We focus specifically on the qubit subspace $\{|00\rangle,|01\rangle,|10\rangle,|11\rangle\}$ of the nonlinear oscillator Hilbert space and assume that the drive is sufficiently weak and off-resonant such that the dispersive approximation holds (see App.~\ref{sec:twoosc}). The coherent state response of an initially unpopulated resonator in the rotating frame of the drive to a pulse with complex envelope $\tilde{\epsilon}(t)$ for state $|jk\rangle$ is
\begin{equation}
\begin{aligned}
\alpha_{jk}^{(0)}(t) & = -\frac{i}{2}\int_0^t e^{-\tilde{\Delta}_{jk}(t-t')}\tilde{\epsilon}(t')dt' \\
& = -\frac{i}{2}e^{-\tilde{\Delta}_{jk}t}\int_0^t e^{\tilde{\Delta}_{jk}t'}\tilde{\epsilon}(t')dt'
\end{aligned}
\end{equation}
where $\tilde{\Delta}_{jk}:=-i(\Delta+\bar{\chi}_{jk})+\kappa/2$, the superscript $(0)$ denotes the initially unpopulated resonator state, and $jk\in\{00,01,10,11\}$. The complex envelope is a phasor representation of the in-phase quadrature (IQ)-modulated tone. The drive frequency $\omega_d$ is detuned by $\Delta$ above the ground state resonator frequency $\omega_r+\chi_{00}$, i.e. $\omega_d=\omega_r+\chi_{00}+\Delta$. This choice allows appreciable drive detuning from all of the dressed resonator peaks and nonlinear oscillator frequencies. For state $|jk\rangle$, $\omega_r+\chi_{jk}$ is the corresponding resonator frequency and $\bar{\chi}_{jk}=\chi_{00}-\chi_{jk}$ is the resonator shift (see Fig.~\ref{fig:peaks}). Finally, $\kappa$ is the photon loss rate of the resonator. We do not include phenomenological amplitude ($T_1$) and phase ($T_2$) damping in the model. Let $jk,lm\in\{00,01,10,11\}$ index the basis states of the qubit pair. The matrix element $|jk\rangle\langle lm|$ of the qubits' reduced density matrix acquires a complex phase factor $e^{i\mu_{jk,lm}(t)}$ where
\begin{equation}
\begin{aligned}
\mu_{jk,lm}(t) & = \mu_{jk,lm}(0) \\
& + (\bar{\chi}_{jk}-\bar{\chi}_{lm})\int_0^t\left[\alpha_{lm}^{(0)}(t')\right]^\ast\alpha_{jk}^{(0)}(t')dt'\\
& + \zeta_0((-1)^{l+m}-(-1)^{j+k})t/4\label{eq:phaseevol}
\end{aligned}
\end{equation}
and $\zeta_0$ is a constant phase accumulation rate due to the static coupling (see App.~\ref{sec:alwaysZZ}) \cite{dicarlo09}. The solution (see App.~\ref{sec:solution}) generalizes naturally to larger numbers of coupled nonlinear oscillators.

When the drive adiabatically traverses a closed path in the complex plane, the phase can be expressed as the sum of dynamical and geometric components \cite{bohm03}. The geometric phase has been studied theoretically and experimentally for a superconducting qubit coupled to a resonator \cite{pechal12,pechal11}. Here we do not consistently require drive shapes to be slowly varying or to enclose non-zero area, but it can be shown that Eq.~\ref{eq:phaseevol} contains the dynamical and geometric contributions to the phase (see App.~\ref{sec:geometric}).

Let ${\cal E}_t(\rho)$ be the quantum operation on the qubit subsystem after evolving with the applied drive for time $t$. From Eq.~\ref{eq:phaseevol}, ${\cal E}_t(|jk\rangle\langle jk|)=|jk\rangle\langle jk|$ and ${\cal E}_t(|jk\rangle\langle lm|)=e^{i(\mu_{jk,lm}(t)-\mu_{jk,lm}(0))}|jk\rangle\langle lm|$. A diagonal two qubit unitary evolution $\mathrm{diag}(e^{i\theta_{00}},e^{i\theta_{01}},e^{i\theta_{10}},e^{i\theta_{11}})$ transforms each matrix element $\rho_{jk,lm}$ of an input density matrix \footnote{The complex number $\rho_{jk,lm}$ denotes the matrix element associated to $|jk\rangle\langle lm|$.} to $e^{i(\theta_{jk}-\theta_{lm})}\rho_{jk,lm}$. Comparing this to ${\cal E}_t(|jk\rangle\langle lm|)$ gives the correspondence $\mu_{jk,lm}(t)=\theta_{jk}-\theta_{lm}$. Therefore, if ${\cal E}_t(\rho)$ is a diagonal unitary evolution then $\mu_{ij,kl}(t)+\mu_{kl,mn}(t)=\mu_{ij,mn}(t)$ and each $\mu_{ij,kl}(t)$ is a real number. This equation is satisfied in the limit of very low loss ($\kappa\ll 1$) for the steady state response, as can be verified by direct calculation from Eq.~\ref{eq:sslowlossphase} in Sec.~\ref{sec:constant}. When $\kappa>0$, it is evident that ${\cal E}_t(\rho)$ dephases in the dressed basis but does not produce unwanted transitions, i.e. leakage errors.

An entangling gate ${\cal E}(\rho)$ can be realized by choosing pulses of appropriate duration, amplitude, and shape. A pulse sequence of total duration $t_g$ realizes an approximately unitary evolution up to a global phase, provided that $\alpha_{jk}(t_g)\approx 0$ and $\mathrm{Im}[\mu_{jk,lm}(t_g)]\approx 0$ for all $jk, lm$. Here $Z=|0\rangle\langle 0|-|1\rangle\langle 1|$ and $X=|0\rangle\langle 1|+|1\rangle\langle 0|$ denote Pauli matrices. The square root of a controlled-Z gate is realized up to single qubit phase rotations by choosing
\begin{equation}
\theta := \mathrm{Re}\left(\mu_{00,01}(t_g)+\mu_{00,10}(t_g)-\mu_{00,11}(t_g)\right)\equiv\pi/2.
\end{equation}
In this paper, we implement a full controlled-Z gate and eliminate the single qubit rotations by applying the composite pulse sequence ${\cal F}\circ {\cal E}\circ {\cal F}\circ {\cal E}$ where ${\cal F}(\rho):=(X\otimes X)\rho(X\otimes X)$ and ${\cal E}$ is the square root of a controlled-Z gate. This corresponds to physically implementing an echo sequence. For simplicity, we assume that ${\cal F}(\rho)$ is applied without error by instantaneous gates. In practice, single qubit gates are faster and have significantly higher fidelity than two qubit gates.

\section{Response to unmodulated tone}\label{sec:constant}

First we quantify the rate of phase accumulation and dephasing for one of the simplest possible cases.
When a tone of constant amplitude is applied to the bus resonator, the resonator reaches its steady state when $t\gg 1/\kappa$. Assuming $\tilde{\epsilon}(t)=\tilde{\epsilon}_0$ and $\kappa>0$, the bus steady state response is
$\alpha_{jk}^{\mathrm{s.s}} = -\frac{i\tilde{\epsilon}_0}{2\tilde{\Delta}_{jk}}$.
At the steady state, the complex phase evolves at a time-independent rate
\begin{equation}
\begin{aligned}
\dot{\mu}_{jk,lm}^{\mathrm{s.s.}} & = (\bar{\chi}_{jk}-\bar{\chi}_{lm})\frac{|\tilde{\epsilon}_0|^2}{4\tilde{\Delta}_{lm}^\ast\tilde{\Delta}_{jk}} \\
& + \zeta_0((-1)^{l+m}-(-1)^{j+k})/4
\end{aligned}
\end{equation}
where $\zeta_0$ is the rate of residual (always-on) $ZZ$ interaction (see App.~\ref{sec:alwaysZZ}).

In the limit of very low loss ($\kappa\ll 1$) and neglecting the static coupling, the real phase advances at a rate
\begin{equation}\label{eq:sslowlossphase}
\mathrm{Re}[\dot{\mu}_{jk,lm}^{\mathrm{s.s.}}] \approx \frac{(\bar{\chi}_{jk}-\bar{\chi}_{lm})|\tilde{\epsilon}_0|^2}{4(\Delta+\bar{\chi}_{jk})(\Delta+\bar{\chi}_{lm})}
\end{equation}
and coherence is lost at a rate
\begin{equation}
\mathrm{Im}[\dot{\mu}_{jk,lm}^{\mathrm{s.s}}] \approx \frac{(\bar{\chi}_{jk}-\bar{\chi}_{lm})^2|\tilde{\epsilon}_0|^2\kappa}{8(\Delta+\bar{\chi}_{jk})^2(\Delta+\bar{\chi}_{lm})^2},
\end{equation}
i.e. the off-diagonal matrix elements of the density matrix decay as $e^{-\mathrm{Im}[\dot{\mu}_{jk,lm}^{\mathrm{s.s.}}]t}$. 

A reasonable approximation for nearly identical qubit-resonator couplings, frequencies, and anharmonicities is $\bar{\chi}:=\bar{\chi}_{01}=\bar{\chi}_{10}=\bar{\chi}_{11}/2$. In this case, phase advances at a rate
\begin{equation}
\dot{\theta}^{\mathrm{s.s.}} \approx -\frac{|\tilde{\epsilon}_0|^2\bar{\chi}^2}{2\Delta(\Delta+\bar{\chi})(\Delta+2\bar{\chi})}-\zeta_0
\end{equation}
and the fastest dephasing occurs between the most detuned states $|00\rangle$ and $|11\rangle$ at the rate
\begin{equation}
\begin{aligned}
\mathrm{Im}[\dot{\mu}_{00,11}^{\mathrm{s.s.}}] & \approx \frac{|\tilde{\epsilon}_0|^2\bar{\chi}^2\kappa}{2\Delta^2(\Delta+2\bar{\chi})^2} \\
& \approx -(\dot{\theta}^{\mathrm{s.s.}}+\zeta_0)\frac{\Delta+\bar{\chi}}{\Delta(\Delta+2\bar{\chi})}\kappa.
\end{aligned}
\end{equation}
These concise expressions show that the entangling rate at the steady state is proportional to the drive power, quadratic in the resonator shift, and suppressed by roughly $\Delta^3$, i.e. ``$\epsilon^2\chi^2/\Delta^3 - \zeta_0$''. The dephasing rate is proportional to the magnitude of the entangling rate times roughly $\kappa/\Delta$, i.e. ``$(\dot{\theta}+\zeta_0)\kappa/\Delta$''. If we assume a fixed photon loss rate $\kappa$, a reasonable strategy is to choose an entangling rate and increase the detuning until the dephasing rate is sufficiently suppressed. As the detuning increases, the drive power must increase like the cube of the detuning to maintain the entangling rate.

\begin{figure}
\centering
{
  \includegraphics[width=.4\textwidth]{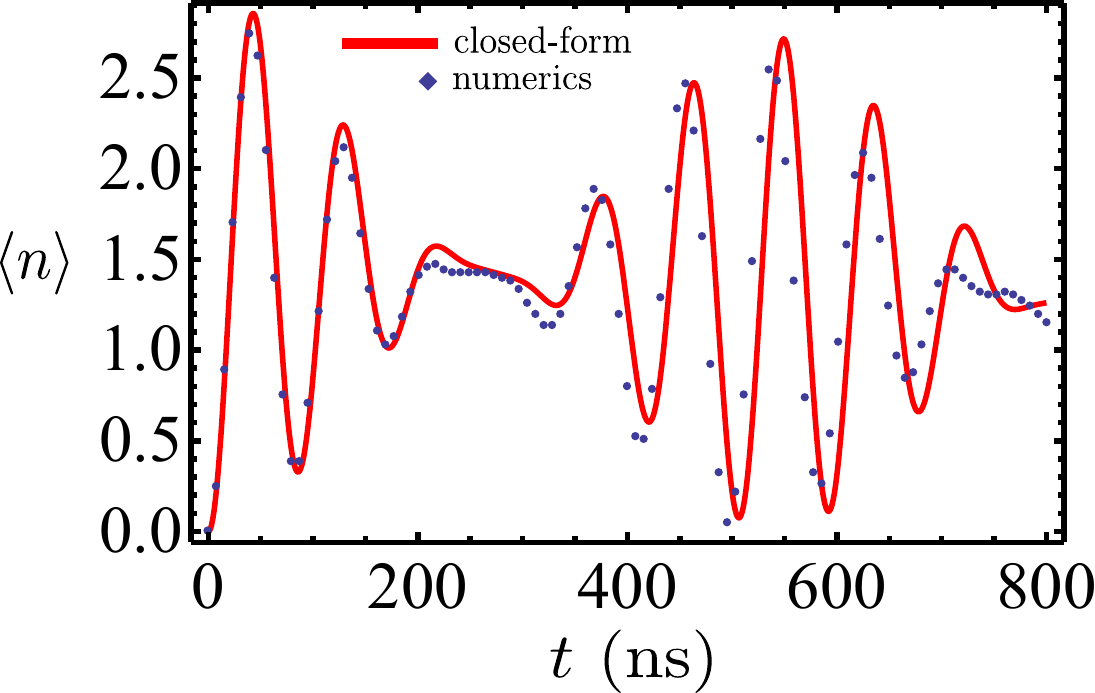}
  \caption{(color online) The bus resonator is driven by an unmodulated tone as described in the text. The mean photon number $\langle n\rangle$ of the bus is calculated from closed-form expressions and by direct numerical solution of the master equation. Both solutions are consistent with one another on this time scale.\label{fig:nbar}}
}
\par \medskip \vfill
{
  \includegraphics[width=.4\textwidth]{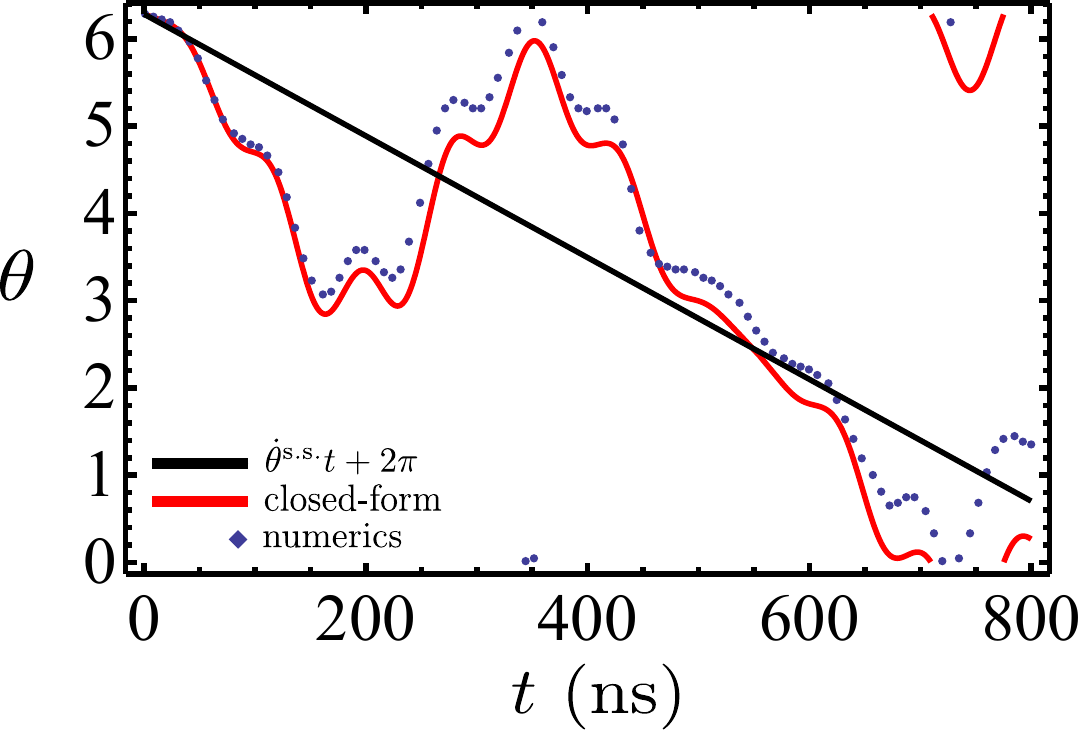}
  \caption{(color online) A pair of qubits coupled to a driven bus resonator accumulate phase as described in the text. The closed-form and numerical solutions for the phase angle $\theta(t)$ are consistent over the time scale shown. The solid line shows a linear accumulation of phase at the steady state rate.\label{fig:theta}}
}
\par \medskip \vfill
{
  \includegraphics[width=.4\textwidth]{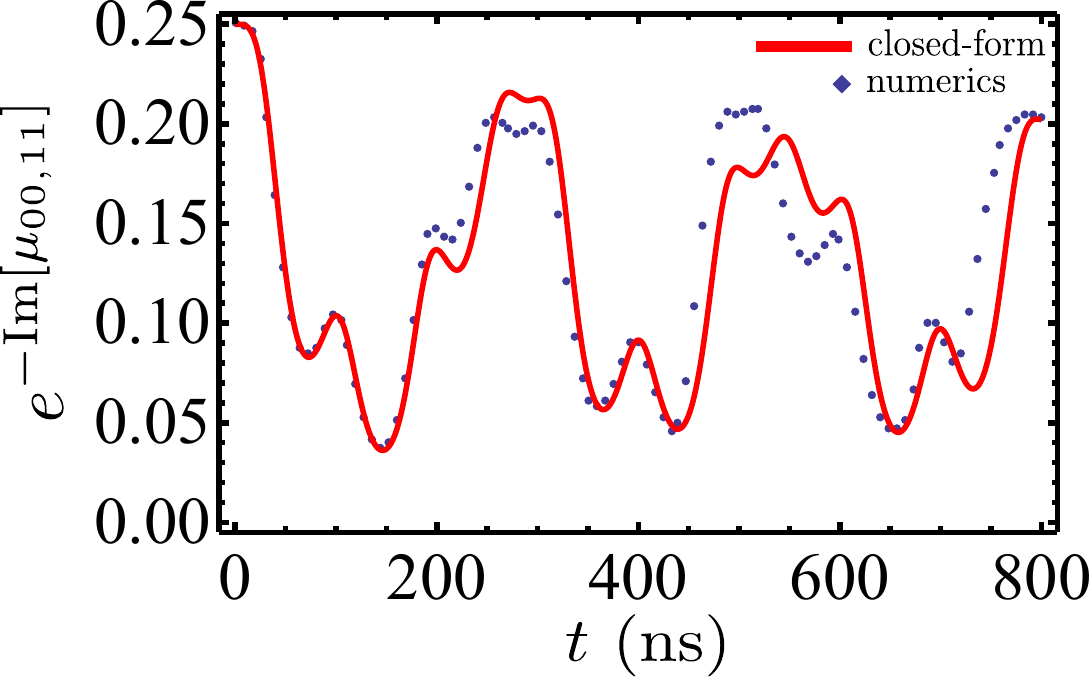}
  \caption{(color online) A pair of qubits coupled to a driven bus resonator suffer measurement-induced dephasing as described in the text. The time evolution of the squared off-diagonal matrix element between $|00\rangle$ and $|11\rangle$ decays exponentially to zero over many microseconds $t\gg 1/\kappa$. The closed-form and numerical solutions are consistent over the time scale shown.\label{fig:immu}}
}
\end{figure}

The closed form solution and its steady state behavior for a sufficiently off-resonant drive are consistent with direct numerical solution of the master equation with $15$ resonator levels (see App.~\ref{sec:numerics}). The initial state of the qubits is chosen to be $|++\rangle\propto |00\rangle+|01\rangle+|10\rangle+|11\rangle$ in the dressed basis, and the bus resonator begins in its ground state. A tone with constant amplitude $|\epsilon/2\pi|=20$\ MHz, fixed phase, and $\Delta/2\pi=10$\ MHz detuning drives the bus resonator. Appendix~\ref{sec:param} lists the remaining parameters of the system ($g_j/2\pi=120\ \mathrm{MHz}$, $\delta_j/2\pi=-300\ \mathrm{MHz}$, $\omega_r/2\pi=7\ \mathrm{GHz}$) using the ``low'' set of qubit frequencies ($\omega_1/2\pi=5\ \mathrm{GHz}$, $\omega_2/2\pi=4.95\ \mathrm{GHz}$). Fig.~\ref{fig:nbar} is a plot of the mean photon number $\langle n\rangle=\mathrm{Tr}\left[\hat{n}\rho(t)\right]$ and shows that the numerical and analytical solutions agree closely over an $800$ ns time interval. This is not sufficient for the system to reach its equilibrium value of $\langle n\rangle\approx 0.7$. Fig.~\ref{fig:theta} is a plot of the numerical and analytically derived phase angle $\theta(t)$ together with the steady state phase accumulation rate. The transient response of the system can be seen to oscillate about the asymptotic approximation $\dot{\theta}^{\mathrm{s.s.}}t+2\pi$. Finally, Fig.~\ref{fig:immu} is a plot of measurement-induced dephasing $\mathrm{exp}\left(-\mathrm{Im}[\mu_{00,11}(t)]\right)$ which again shows consistency between numerics and closed-form solutions. The off-diagonal matrix element begins at $1/4$ as expected and decays slowly but exponentially toward zero while oscillating about the asymptotic approximation $\mathrm{exp}(-\mathrm{Im}[\dot{\mu}_{00,11}^{\mathrm{s.s.}}]t)/4$. Deviations from the asymptotic approximations become negligible for $t>2\pi/\kappa=20$ $\mu$s for all of these quantities.

\section{Spline shaped pulse}\label{sec:spline}

A constant tone entangles qubits at a rate that is well understood, but it leaves the resonator in an excited state that depends on the qubit state. If the resonator is not disentangled from the qubits, the residual entanglement leads to additional decoherence. One approach to solving this problem is to attempt to leave the resonator in the ground state after the gate is complete.

We begin by integrating the resonator response by parts
\begin{equation}
\begin{aligned}
\int_0^{t} e^{\tilde{\Delta}_{jk}t'}\tilde{\epsilon}(t')dt' & = \frac{\tilde{\epsilon}(t)e^{\tilde{\Delta}_{jk}t}}{\tilde{\Delta}_{jk}}-\frac{\tilde{\epsilon}(0)}{\tilde{\Delta}_{jk}}\\
& -\frac{1}{\tilde{\Delta}_{jk}}\int_0^te^{\tilde{\Delta}_{jk}t'}\dot{\tilde{\epsilon}}(t')dt'.
\end{aligned}
\end{equation}
Iterating the integral $M$ times, we arrive at the expansion
\begin{equation}\label{eq:adiabatic}
\begin{aligned}
\alpha_{jk}^{(0)}(t) & =\frac{ie^{-\tilde{\Delta}_{jk}t}}{2}\sum_{m=0}^{M-1}(-1)^{m}\tilde{\epsilon}^{(m)}(0)/\tilde{\Delta}_{jk}^{m+1} \\
& -\frac{i}{2}\sum_{m=0}^{M-1} (-1)^m \tilde{\epsilon}^{(m)}(t)/\tilde{\Delta}_{jk}^{m+1} \\
& + (-1)^{M+1}\frac{ie^{-\tilde{\Delta}_{jk}t}}{2\tilde{\Delta}_{jk}^M}\int_0^te^{\tilde{\Delta}_{jk}t'}\tilde{\epsilon}^{(M)}(t')dt'
\end{aligned}
\end{equation}
where $\tilde{\epsilon}^{(m)}(t)$ denotes the $m$th derivative of $\tilde{\epsilon}(t)$.
A pulse shape of duration $t_g$ that satisfies $\tilde{\epsilon}^{(m)}(0)=\tilde{\epsilon}^{(m)}(t_g)=0$ for all $m<M$ will have a residual resonator population that is suppressed by a power of the detuning. 

\begin{figure}
\centering
  \includegraphics[width=.4\textwidth]{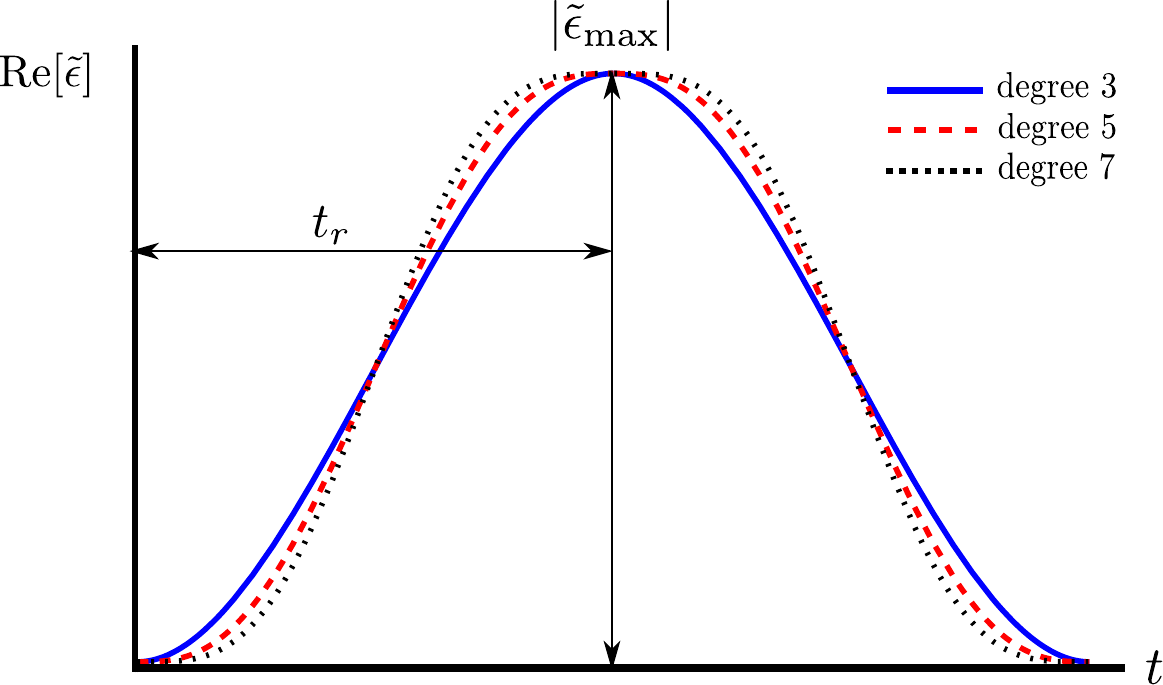}
  \caption{(color online) Pulse shapes constructed from piecewise polynomials rise from amplitude zero to $|\tilde{\epsilon}_\mathrm{max}|$ over time $t_r$ and are symmetric about $t=t_r+t_p/2$. Such pulse shapes have some number of vanishing derivatives at times $t=0$ and $t=2t_r+t_p$. The first three examples are shown here and we take $\mathrm{Im}[\tilde{\epsilon}]=0$ and $t_p=0$.\label{fig:splines}}
\end{figure}

One particularly simple solution is to use piecewise polynomial shapes of a given degree that satisfy certain smoothness criteria, i.e. splines. We consider a scaled symmetric pulse $\tilde{\epsilon}_0p_d(t)$ of the form
\begin{equation}
p_d(t) := \left\{\begin{array}{cc}
s_d(t,t_r), & t\in [0,t_r] \\
1, & t\in (t_r,t_r+t_p] \\
s_d(-t+2t_r+t_p,t_r), & t\in (t_r+t_p,\\
& 2t_r+t_p]
\end{array}\right.
\end{equation}
where the odd degree $d$ polynomials $s_d(t,t_r)$ satisfy $s_d(0,t_r)=0$ and $s_d(t_r,t_r)=1$, $t_r>0$ is the rise time of the pulse, and the middle of the pulse has duration $t_p\geq 0$.
Consider polynomials of the form
\begin{equation}
s_d(t,t_r) := t^{(d+1)/2}\sum_{m=0}^{(d-1)/2} a_mt^m
\end{equation}
where $a_m:=(-1)^mc_m/t_r^{(d+1)/2+m}$ are scaled coefficients derived from positive integers $\{c_0,c_1,\dots,c_m,\dots,c_{(d-1)/2}\}$.
It is possible to find polynomials $s_d$ whose $j$th derivatives vanish at the boundary of the interval $[0,t_r]$ for all $1\leq j\leq (d-1)/2$ by solving
the system of linear equations
\begin{equation}\label{eq:spline}
\sum_{m=0}^{(d-1)/2}(-1)^mc_m{(d+1)/2+m\choose j}j!  = \delta(j),
\end{equation}
where $j=0,1,\dots,(d-1)/2$ and $\delta(j)$ takes the value $1$ if $j=0$ and $0$ otherwise. The maximum absolute value of the lowest non-vanishing derivative of $s_d$ occurs on the boundary of the interval $[0,t_r]$ and equals $c_0((d+1)/2)!/t_r^{(d+1)/2}$. The first several solutions are given in Table~\ref{tab:sol} and shown in Fig.~\ref{fig:splines}.

\begin{table}
\begin{tabular}{c|c|c}
$d$ & $\{c_0,c_1,\dots\}$ & $|\mathrm{max}\frac{\partial^{(d-1)/2+1} s}{\partial t^{(d-1)/2+1}}|$ \\ \hline
$3$ & $\{3,2\}$ & $6/t_r^2$ \\
$5$ & $\{10,15,6\}$ & $60/t_r^3$ \\
$7$ & $\{35,84,70,20\}$ & $840/t_r^4$ \\
$9$ & $\{126, 420, 540, 315, 70\}$ & $15120/t_r^5$ \\
$11$ & $\{462, 1980, 3465, 3080, 1386, 252\}$ & $332640/t_r^6$
\end{tabular}
\caption{Solutions to Eq.~\ref{eq:spline} for the first several odd degrees $d$ and the maximum magnitude of the first non-vanishing derivative\label{tab:sol}}
\end{table}

Applying the Schwartz inequality to the tail of Eq.~\ref{eq:adiabatic}, the expected number of resonator photons at the end of a pulse is bounded by
\begin{align}
\langle n_{jk}^{(0)}(t_g)\rangle & = \frac{e^{-\kappa t_g}}{4|\tilde{\Delta}_{jk}|^{2M}}\left| \int_0^{t_g} e^{\tilde{\Delta}_{jk}t'}\tilde{\epsilon}^{(M)}(t')dt' \right|^2 \label{eq:residual1}\\
& \leq |\tilde{\epsilon}^{(M)}_\mathrm{max}|^2t_g\frac{1-e^{-\kappa t_g}}{4\kappa|\tilde{\Delta}_{jk}|^{2M}}\label{eq:residual}
\end{align}
where $|\tilde{\epsilon}^{(M)}_\mathrm{max}|$ is the largest absolute value of the $M^\mathrm{th}$ derivative of the envelope on $[0,t_g]$. Assuming a constant value of $\kappa$, this general bound shows that it is sufficient for the detuning to be significantly greater than $\sqrt[2M]{|\tilde{\epsilon}^{(M)}_\mathrm{max}|^2t_g}$, but the bound is weak. For the polynomial shapes
\begin{equation}
\langle n_{jk}^{(0)}(t_g)\rangle \leq \frac{[c_0((d+1)/2)!]^2t_g}{4\kappa(t_r|\tilde{\Delta}_{jk}|)^{d+1}}(1-e^{-\kappa t_g}).\label{eq:weak}
\end{equation}

\begin{figure}
\centering
{
  \includegraphics[width=.4\textwidth]{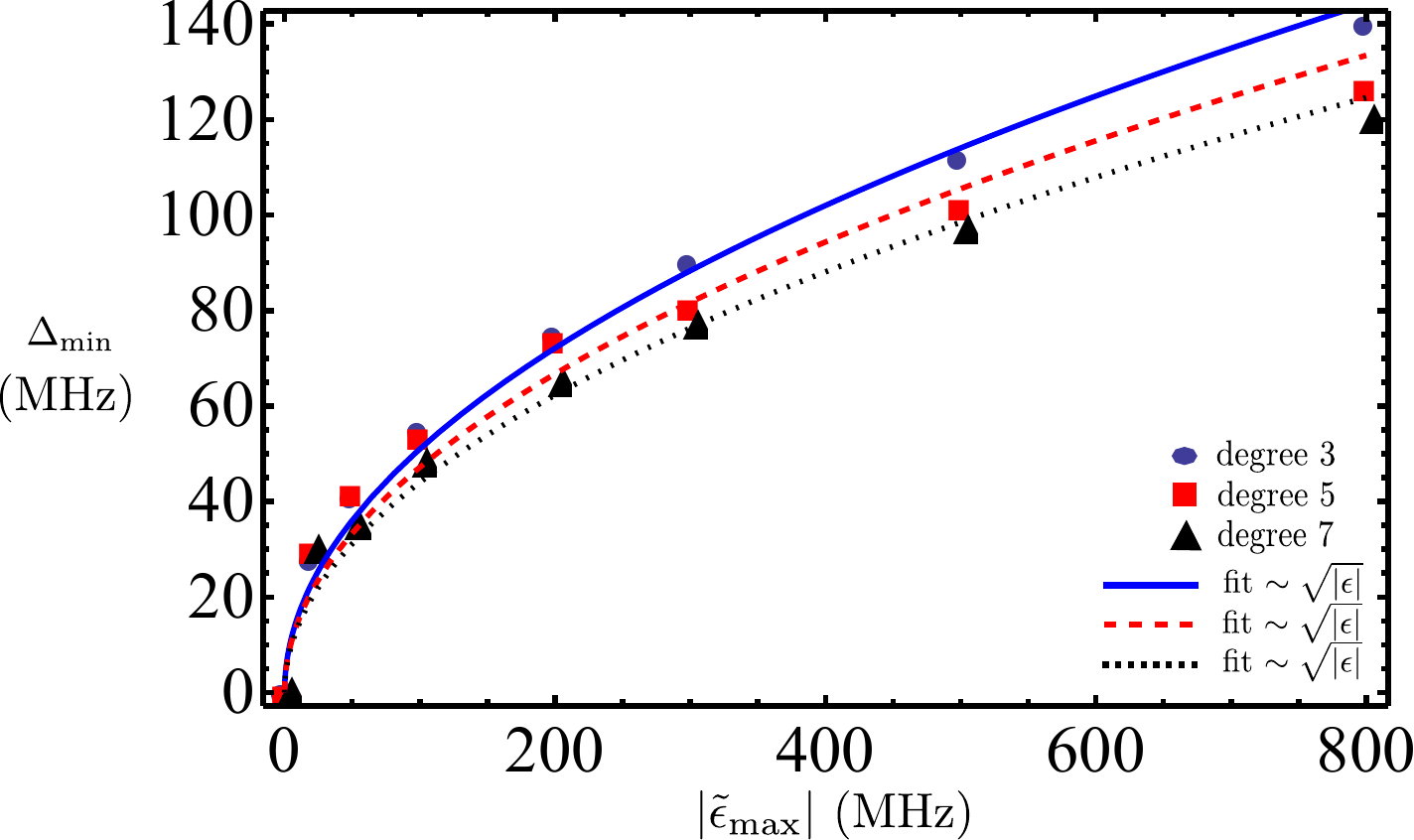}
  \caption{(color online) This plot of the drive detuning versus the peak drive amplitude (as a Rabi rate) shows roughly a square root dependence of the detuning on the drive amplitude for splines of degree 3, 5, and 7. The composite gates in each case have pulse shapes that achieve $10^{-4}$ average gate infidelity. The corresponding curves are fits to the form $c\sqrt{|\tilde\epsilon|}$ where $c$ is a fitting parameter.\label{fig:detuningspline}}
}
\par \medskip \vfill
{
  \includegraphics[width=.4\textwidth]{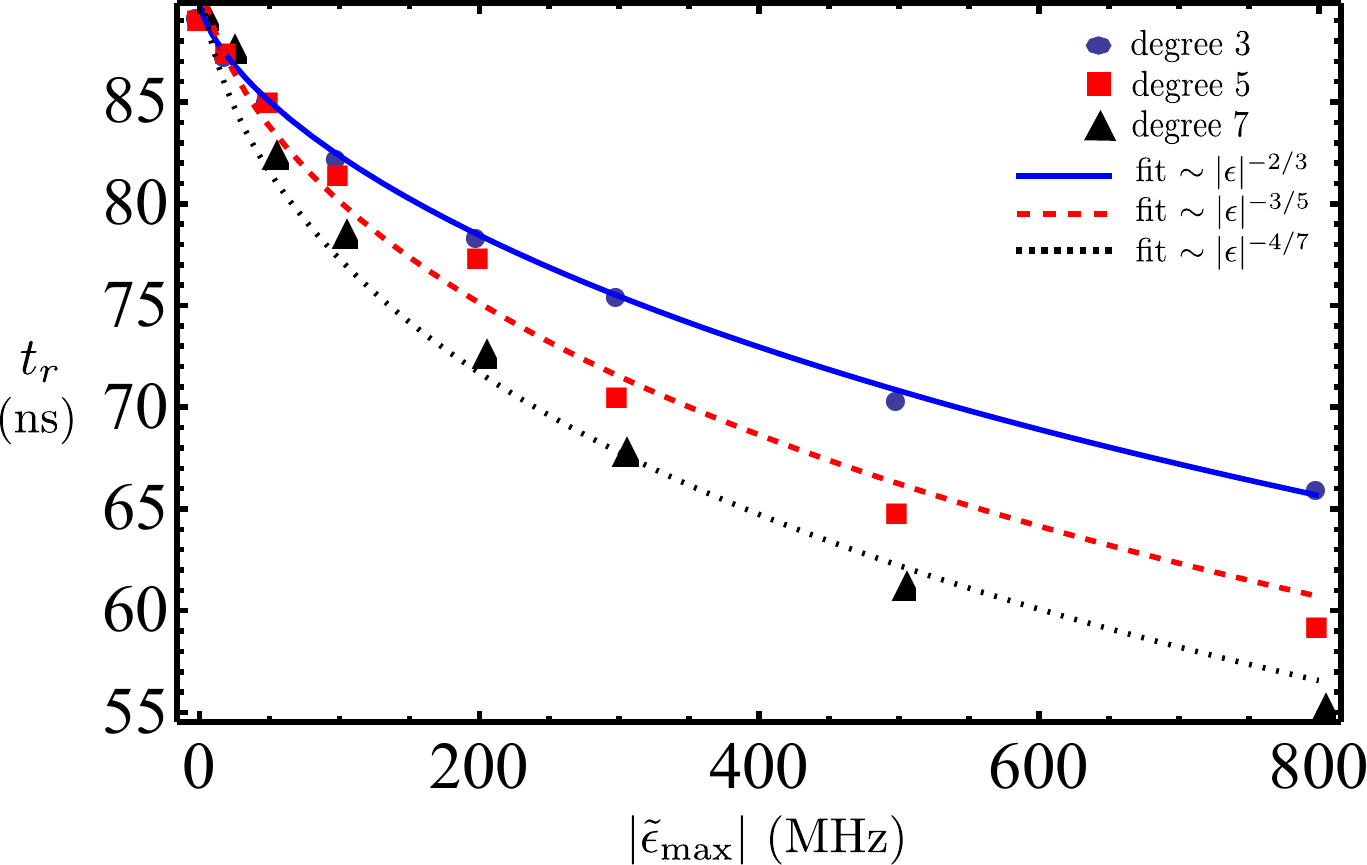}
  \caption{(color online) For the same set of pulse shapes as Figure~\ref{fig:detuningspline}, this plot shows the rise time of each pulse in the composite sequence as a function of drive amplitude. The corresponding curves are fits to the functional form $c_0/(c_1+c_2|\tilde{\epsilon}|^{(d+1)/(2d)})$ where $c_j$ are fitting parameters that depend on $d$.\label{fig:timespline}}
}
\end{figure}

The piecewise polynomial shapes can be exactly integrated using the easily derived identity
\begin{equation}
\int e^{cx}x^k dx = \sum_{m=0}^k (-1)^m\frac{k!}{(k-m)!}\frac{e^{cx}}{c^{m+1}}x^{k-m}
\end{equation}
where $k$ is a non-negative integer and $c$ is a complex number. Because the solutions have a closed form, the space of gates with spline pulses can be rapidly explored. From this point on, we set $t_p=0$ everywhere. For a fixed set of system parameters and spline pulse degree, the average gate infidelity is a function of the peak drive amplitude $|\tilde{\epsilon}_\mathrm{max}|$, its detuning $\Delta$, and the rise time $t_r$ of the spline pulse. To achieve some target infidelity, we use the following procedure. At each peak drive amplitude and detuning, we choose the shortest pulse rise time $t_r(|\tilde{\epsilon}|_\mathrm{max},\Delta)>0$ that maximizes the average gate fidelity to a controlled-Z gate, realized as a composite pulse using a pair of spline pulses. Recall that the only dissipation in our model comes about from the lossy resonator, so at each peak drive amplitude the fidelity tends to increase as the detuning increases. Therefore at given peak drive amplitude, we increase the detuning until reaching a value $\Delta_\mathrm{min}>0$ such that the controlled-Z gate achieves the target infidelity. In practice, the gate can be tuned similarly and hence is robust to imprecise knowledge of the resonator shifts. We carry out this procedure for each peak drive amplitude to study the dependence of $\Delta_\mathrm{min}$ and $t_r$ on the peak drive amplitude for a fixed target infidelity.

Motivated by the goal of achieving a gate whose error rate is below an accuracy threshold for fault-tolerant quantum computing, we set a target average gate infidelity of $10^{-4}$. Taking the ``high'' set of system parameters ($\omega_1/2\pi=5.6\ \mathrm{GHz}$, $\omega_2/2\pi=5.55\ \mathrm{GHz}$, see App.~\ref{sec:param}), all of the resulting high-fidelity gates inject at most $5$ resonator photons at peak. Fig.~\ref{fig:detuningspline} is a plot of $\Delta_\mathrm{min}/2\pi$ as a function of the peak drive amplitude $|\tilde{\epsilon}_\mathrm{max}/2\pi|$ for spline pulse shapes of degrees $3$, $5$, and $7$. Based on the approximate scaling obtained in Sec.~\ref{sec:constant}, we expect $|\tilde{\epsilon}_\mathrm{max}|^2\sim \Delta_\mathrm{min}^4$ if $\mathrm{Im}[\dot{\mu}_{00,11}]$ is held constant. The results show that this is approximately valid for each of the splines shown.

Fig.~\ref{fig:timespline} is a plot of the minimum pulse rise time $t_r$ as a function of the peak drive amplitude for the same sequence of splines. The drive detuning at each amplitude is given by the corresponding point of Fig.~\ref{fig:detuningspline}. At low drive powers, the gate rate is dominated by the static coupling, so the drive power must be increased to enter a regime where the phase accumulation rate has a significant drive-induced component. In this regime, increasing the degree of the spline reduces the rise time of the pulse, leading to a faster controlled-Z gate. Although it is not shown here, this trend must reverse eventually due to the $M$th derivative in Eq.~\ref{eq:residual1}, and indeed we have confirmed that increasing the degree to $d>7$ does not further reduce the rise time for this example.

To get a rough idea of the functional form of $t_r$, suppose the infidelity is due entirely to residual bus photons, and therefore the fidelity target corresponds to holding $\langle n_{jk}^{(0)}(t_g)\rangle$ constant. The exact expression for the number of residual photons involves an integral of the $M$th derivative, but the bound of Eq.~\ref{eq:weak} is too weak to show a benefit to increasing the degree of the spline. However, the bound suggests that $[((d+1)/2)!]^2t_g/(t_r\Delta)^{d+1}$ changes slowly. Neglecting the prefactor $[((d+1)/2)!]^{2}$ and approximating $\Delta\sim\sqrt{\tilde{\epsilon}}$, we arrive at a rough expression $t_r\sim |\tilde{\epsilon}|^{-(d+1)/(2d)}$. In Fig.~\ref{fig:timespline}, we fit results for the optimized gate sequences to the functional form $c_0/(c_1+c_2|\tilde{\epsilon}|^{(d+1)/(2d)})$ as a guide. The fit is poorer if the power of $|\tilde{\epsilon}|$ is taken to be independent of $d$.

For later comparison with numerically optimized pulse shapes, we use the same approach as above to estimate the minimum gate time using splines with parameters $|\tilde{\epsilon}_\mathrm{max}/2\pi|\approx 284$ MHz and $d=7$ with a fidelity goal of $\sim 6\times 10^{-4}$. The spline realizing a square root of controlled-Z gate for this peak drive amplitude has $t_r\approx 53$ ns at a detuning of $\Delta/2\pi\approx 57\ \mathrm{MHz}$, so the composite gate has a $212$ ns duration.

\section{Nullspace Optimization}\label{sec:nullspace}

Spline pulses are simple shapes for implementing high fidelity gates, but operating in the adiabatic limit and using only a single quadrature is unlikely to be optimal. To limit decoherence due to dissipation ($T_1$), there is motivation to consider fast pulses. However, fast pulses can more easily leave the resonator in a populated state, and that state differs for each branch of the wave function, leading to decoherence. One strategy for overcoming this limitation is to apply a train of short pulses such that the resonator is ``reset'' or ``unwound'' at the end of the train, i.e. for all qubit states $|jk\rangle$, $\alpha_{jk}^{(0)}(t_g)\approx 0$. 

We will return to this reset problem shortly and first consider how to express solutions for pulse shapes that are convenient for numerical optimization. One approach is to express the resonator response as a function of a vector of pulse parameters. We consider a discretized pulse shape given by a train of steps of the form
\begin{equation}\label{eq:discretepulse}
\tilde{\epsilon}(t) = \sum_{q=0}^{N-1} \tilde{\epsilon}_q\left[u(t-q\Delta t)-u(t-(q+1)\Delta t)\right]
\end{equation}
where $u(x)$ is the unit step, $\Delta t$ is a time interval, and for each $q$, $\tilde{\epsilon}_q$ is a complex number. The resonator response for state $|jk\rangle$ is
\begin{widetext}
\begin{equation}
\begin{aligned}
\alpha_{jk}(t) & = -\frac{i}{2}e^{-\tilde{\Delta}_{jk}t}\left[ c\left(\frac{t}{\Delta t},N_t\right)\frac{\tilde{\epsilon}_{N_t}}{\tilde{\Delta}_{jk}}(e^{\tilde{\Delta}_{jk}t}-e^{\tilde{\Delta}_{jk}N_t\Delta t}) + \frac{e^{\tilde{\Delta}_{jk}\Delta t}-1}{\tilde{\Delta}_{jk}}\sum_{q=0}^{N_t-1}\tilde{\epsilon}_qe^{\tilde{\Delta}_{jk}q\Delta t}\right]
\end{aligned}
\end{equation}
\end{widetext}
where $N_t=\lfloor t/\Delta t\rfloor$ and $c(x,y)$ equals $0$ if $x=y$ and $1$ otherwise. The final resonator state is given by
\begin{equation}\label{eq:cavfinal}
\alpha_{jk}(N\Delta t) = -\frac{i}{2}\left(\frac{e^{\tilde{\Delta}_{jk}\Delta t}-1}{\tilde{\Delta}_{jk}}\right)e^{-\tilde{\Delta}_{jk}N\Delta t}\sum_{q=0}^{N-1}\tilde{\epsilon}_qe^{\tilde{\Delta}_{jk}q\Delta t}.
\end{equation}

To evaluate the complex phase, we begin by defining
\begin{equation}
I(M\Delta t):=\int_0^{M\Delta t} \left[\alpha_{lm}^{(0)}(t')\right]^\ast\alpha_{jk}^{(0)}(t')dt'
\end{equation}
and write
\begin{equation}
\begin{aligned}
\mu_{jk,lm}(M\Delta t) & =(\bar{\chi}_{jk}-\bar{\chi}_{lm})I(M\Delta t) \\
& +\zeta_0((-1)^{l+m}-(-1)^{j+k})M\Delta t.
\end{aligned}
\end{equation}
This can be evaluated straightforwardly once $\alpha_{jk}(N\Delta t)$ has been computed for each $jk$ and $N$. It is useful to define
\begin{equation}
h_\Delta(x) := (e^{\Delta x}-1)/\Delta
\end{equation}
and compute complex constants
\begin{equation}
\begin{aligned}
U_{jk,lm} & := h_{-\tilde{\Delta}_{jk}-\tilde{\Delta}_{lm}^\ast}(\Delta t) \\
U_{jk} & := h_{-\tilde{\Delta}_{jk}}(\Delta t). \\
\end{aligned}
\end{equation}
Now, $I(M\Delta t)$ is the sum of four terms,
\begin{equation}
\begin{aligned}
& I(M\Delta t) = U_{jk,lm}\sum_{p=1}^{M-1}\left[\alpha_{lm}^{(0)}(p\Delta t)\right]^\ast\alpha_{jk}^{(0)}(p\Delta t) \\
& -\frac{i}{2\tilde{\Delta}_{jk}}\left(U_{lm}^\ast-U_{jk,lm}\right)\sum_{p=1}^{M-1}\tilde{\epsilon}_j\left[\alpha_{lm}^{(0)}(p\Delta t)\right]^\ast \\
& +\frac{i}{2\tilde{\Delta}_{lm}^\ast}\left(U_{jk}-U_{jk,lm}\right)\sum_{p=1}^{M-1}\tilde{\epsilon}_j^\ast\alpha_{jk}^{(0)}(p\Delta t) \\
& +\frac{1}{4\tilde{\Delta}_{jk}\tilde{\Delta}_{lm}^\ast}\left(\Delta t-U_{jk}-U_{lm}^\ast+U_{jk,lm}\right)\sum_{p=0}^{M-1}|\tilde{\epsilon}_j|^2.
\end{aligned}
\end{equation}

Consider a gate implemented by pulses of duration $t_g=M\Delta t$, detuning $\Delta$, and fixed system parameters such as those given in App.~\ref{sec:param}. The first goal is to find a discretized pulse (Eq.~\ref{eq:discretepulse}) such that for all $jk\in\{00,01,10,11\}$ the bus is ``reset'' or ``unwound'' to the vacuum state $\alpha_{jk}(t_g)=0$. This condition is precisely the statement that the length $M$ vector of complex pulse amplitudes $\{\tilde{\epsilon}_q\}$ is an element of the nullspace of
\begin{equation}
\begin{aligned}
A := & \mathrm{diag}\left(h_{\tilde{\Delta}_{00}}(\Delta t)e^{-\tilde{\Delta}_{00}M\Delta t},h_{\tilde{\Delta}_{01}}(\Delta t)e^{-\tilde{\Delta}_{01}M\Delta t},\right.\\
& \left.h_{\tilde{\Delta}_{10}}(\Delta t)e^{-\tilde{\Delta}_{10}M\Delta t},h_{\tilde{\Delta}_{11}}(\Delta t)e^{-\tilde{\Delta}_{11}M\Delta t}\right)\cdot \\
& \left[\begin{array}{ccccc}
1 & e^{\tilde{\Delta}_{00}\Delta t} & e^{\tilde{\Delta}_{00}2\Delta t} & \dots & e^{\tilde{\Delta}_{00}(M-1)\Delta t} \\
1 & e^{\tilde{\Delta}_{01}\Delta t} & e^{\tilde{\Delta}_{01}2\Delta t} & \dots & e^{\tilde{\Delta}_{01}(M-1)\Delta t} \\
1 & e^{\tilde{\Delta}_{10}\Delta t} & e^{\tilde{\Delta}_{10}2\Delta t} & \dots & e^{\tilde{\Delta}_{10}(M-1)\Delta t} \\
1 & e^{\tilde{\Delta}_{11}\Delta t} & e^{\tilde{\Delta}_{11}2\Delta t} & \dots & e^{\tilde{\Delta}_{11}(M-1)\Delta t}
\end{array}
\right].
\end{aligned}
\end{equation}
This matrix is easily calculated and, if $M\geq 4$, the nullspace is non-empty.

The next goal is to obtain a high fidelity gate by optimizing over linear combinations of vectors in the nullspace. Furthermore, we penalize high bandwidth pulses and pulses that are nonzero at $t=0$ and $t=t_g$. Therefore, we choose a cost function of the form
\begin{equation}
\begin{aligned}
C(\vec{\epsilon}) & := \beta_{1}\left(1-{\cal F}_A(U,{\cal G}_{\vec{\epsilon}})\right)
+ \beta_2\sum_k w_k|\tilde{\epsilon}_k|^2 \\
& + \beta_3\sum_{f_k:|f_k|>B} |{\cal F}[\tilde{\epsilon}](f_k)|^2
\end{aligned}
\end{equation}
where $\beta_{1},\beta_{2},\beta_{3}$ are real parameters and ${\cal F}[\tilde{\epsilon}](f_k)$ denotes the discrete Fourier transform of the drive envelope evaluated at a frequency point $f_k$. The $\beta_1$ term ensures that the gate has high average gate fidelity with a controlled-Z gate (see App.~\ref{sec:fidelity}). The $\beta_2$ term enforces a physical constraint that the pulse is ``off'' at the start and end times by minimizing a weighted sum of the power $|\tilde{\epsilon}_k|^2$ with weights $\{w_k\}$ sampled from the function $w(t):=e^{-\gamma t}+e^{\gamma(t-t_g)}$, where we take $\gamma=1\ \mathrm{ns}^{-1}$. The $\beta_3$ term constraints the bandwidth $B$ of the pulse by summing the power outside of a given bandwidth, where we take $B=300\ \mathrm{MHz}$. This choice of objective function has the advantage of correctly weighting angle inaccuracy and measurement-induced dephasing, since each will be reduced as much as is needed to minimize infidelity.

\begin{figure}
\centering
{
  \includegraphics[width=.5\textwidth]{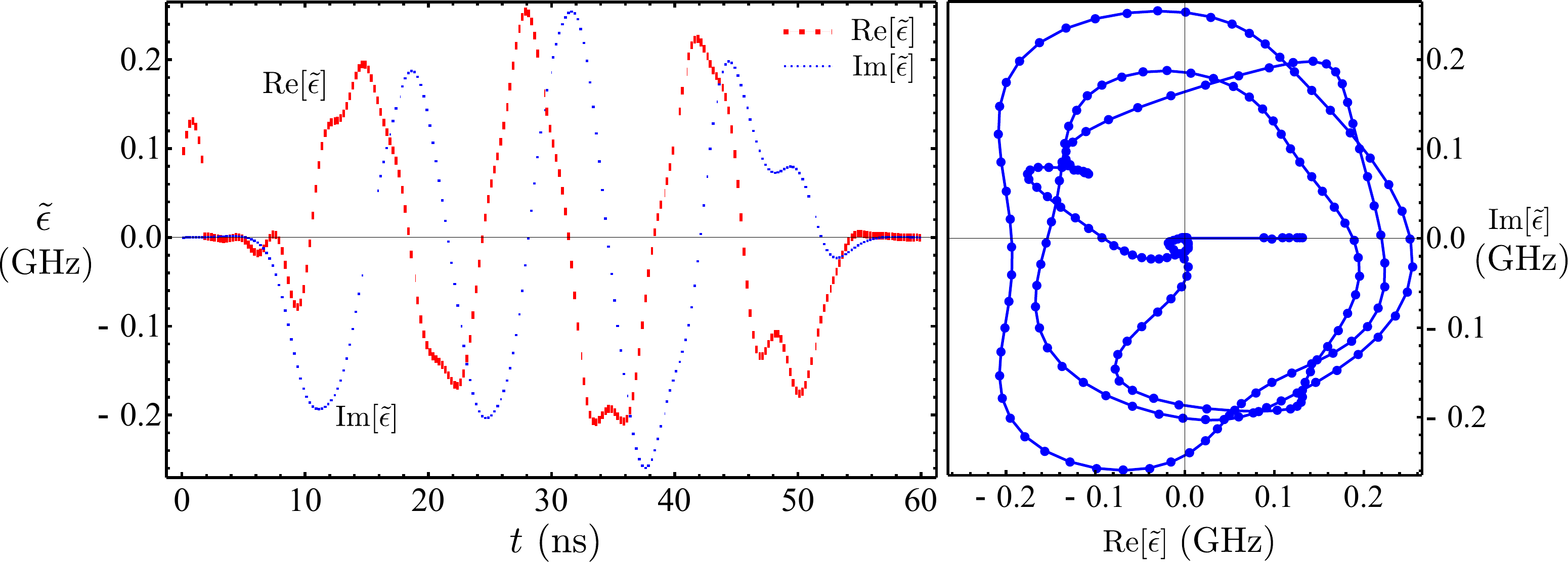}
  \caption{(color online) Example of a pulse shape produced by nullspace optimization. On the left, the in-phase (real) part is plotted in (thick) red and the quadrature (imaginary) part is plotted in (thin) blue. The same pulse shape is plotted on the right in the complex (IQ) plane. System parameters are the ``high'' set given in App.~\ref{sec:param}. The sequence has 240 discrete samples with $\Delta t=0.25$ ns. This example realizes a square root of controlled-Z gate ($\theta=\pi/2$) with angular error $\Delta\theta\approx 2\times 10^{-3}$ and achieves cost components $1-{\cal F}\approx 6\times 10^{-4}$, on-off penalty $\sim 9\times 10^{-5}$, and bandwidth penalty $\sim 5\times 10^{-5}$. \label{fig:optimizationtime}}
}
\end{figure}

\begin{figure}
\centering
{
  \includegraphics[width=.5\textwidth]{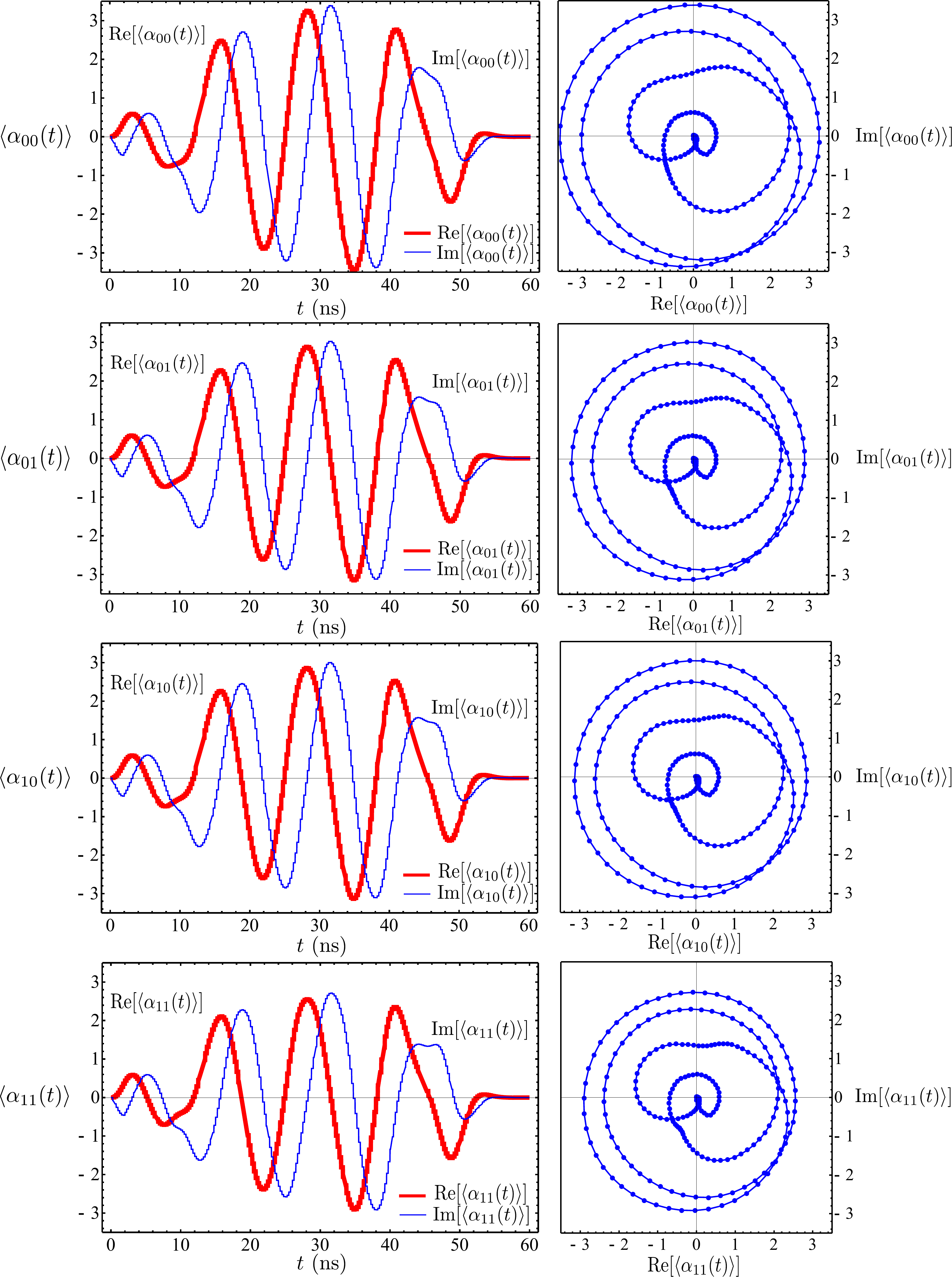}
  \caption{(color online) For each dressed basis state $|jk\rangle$, the pulse shape shown in Fig.~\ref{fig:optimizationtime} causes the bus resonator response $\langle\alpha_{jk}(t)\rangle$ to rapidly ``ring up'', follow a nearly circular trajectory in phase space, and return to the origin. The pulse populates the bus with fewer than $10$ photons at peak for any initial state of the qubits. \label{fig:optimizationalpha}}
}
\end{figure}

\begin{figure}
\centering
{
  \includegraphics[width=.4\textwidth]{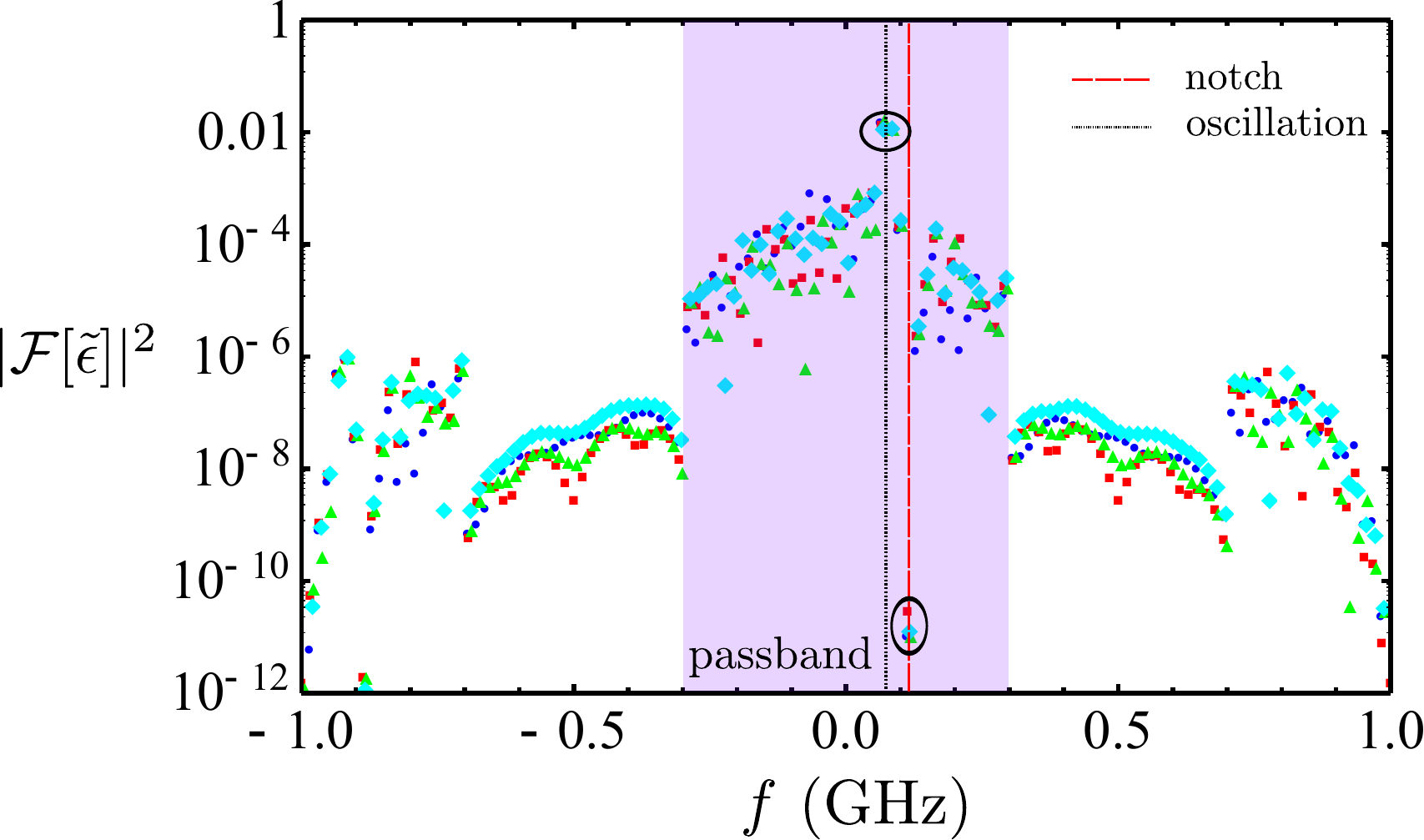}
  \caption{(color online) Spectrum of four different pulses found by nullspace optimization starting from four random bounded initial pulses. Several spectral features (circled) are apparent across all of the solutions. The $B=300$ MHz bandwidth is visible, as is a notch at $\{\Delta+\chi_{jk}\}$ and oscillations at a frequency $\omega/2\pi\approx 60$ MHz. \label{fig:optimizationfft}}
}
\end{figure}

Applying this nullspace method within a gradient descent algorithm, taking the ``high'' set of parameters ($\omega_1/2\pi=5.6\ \mathrm{GHz}$, $\omega_2/2\pi=5.55\ \mathrm{GHz}$) from App.~\ref{sec:param}, and $\Delta/2\pi=112$ MHz, we find pulses that realize a controlled-Z gate with $\sim 99.94$\% fidelity in $\sim 120$ns with a bandwidth of $300$ MHz and peak amplitude $|\tilde{\epsilon}_{\mathrm{max}}/2\pi|\approx 284\ \mathrm{MHz}$ (see Fig.~\ref{fig:optimizationtime}). The corresponding resonator response is shown in Fig.~\ref{fig:optimizationalpha}. The residual cost is dominated by gate infidelity due to measurement-induced dephasing and angular error rather than the penalty functions. One can verify that this solution is not adiabatic in the sense that Eq.~\ref{eq:adiabatic} at $M=2$ poorly approximates the resonator response. The total phase accumulated during the square root of the controlled-Z gate is $\sim -1.569$ radians to which static coupling contributes $\sim -0.527$ radians and the difference is resonator-induced. This resonator-induced phase further separates into a dynamical component of $\sim -0.047$ radians (Eq.~\ref{eq:dynamicphase}), a geometric component of $\sim 4\times 10^{-8}$ radians (Eq.~\ref{eq:geometricphase}), a residual phase of $\sim -0.044$ radians (Eq.~\ref{eq:residualphase}), and the remaining phase is contributed by higher-order terms in the adiabatic expansion.

The solution has clearly discernible features in Fourier space (see Fig.~\ref{fig:optimizationfft}). First, the bandwidth constraint is readily apparent. Second, there is a notch over the range of frequencies $\{\Delta+\chi_{jk}\}$. Third, there is a distinct peak within the passband at around $60$ MHz. These features suggest considering analytical pulse shapes of the form $e^{i\omega_s t}f(t)\ast g(t)$ where $\omega_s$ is some frequency, $f(t)$ is a complex envelope, $g(t)$ is the impulse response of a notch filter, and $\ast$ denotes the convolution product. Considering such shapes in the future may lead to a deeper understanding of how to improve and simplify non-adiabatic dynamical phase gates.

\section{Conclusion}

We have explored theoretical drive pulse shapes for the resonator-induced phase gate, which is a potentially high-fidelity, low-leakage, controlled-Phase gate for superconducting qubits. These pulse shapes minimize gate error under physically reasonable assumptions. Relatively simple close-form solutions enable us to propose and rapidly explore spline pulse shapes that reduce residual qubit-bus entanglement, are robust to imprecise knowledge of the resonator shift, and can be shortened by using higher-degree polynomials. Their simplicity may ease experimental calibration and implementation of high-fidelity phase gates. Finally, to pursue faster gates, we present a new procedure that optimizes over the subspace of pulses that leave the resonator unpopulated. This procedure finds numerical shaped drive pulses that further reduce the gate duration. The shapes revealed by this nullspace optimization have structure that merits further study and could lead to improved analytically-derived shapes.

\begin{acknowledgements}
We thank Lev Bishop, Jerry Chow, Oliver Dial, Stefan Filipp, Hanhee Paik, Stefano Poletto, and Matthias Steffen for helpful comments. We acknowledge support from IARPA under contract W911NF-10-1-0324.
\end{acknowledgements}

\appendix

\section{Two non-linear oscillators coupled to a resonator}\label{sec:twoosc}

A Duffing oscillator with frequency $\omega$, anharmonicity $\delta$, and lowering operator $a$ is described by the Hamiltonian
\begin{equation}
H(\omega,\delta,a)/\hbar=\omega a^\dagger a + \frac{\delta}{2}a^\dagger a(a^\dagger a-1).
\end{equation}
Consider two such oscillators indexed by $\iota\in\{1,2\}$ with frequencies $\omega_\iota$ and anharmonicities $\delta_\iota$ coupled to a bus resonator with frequency $\omega_r$ and coupling strengths $g_\iota$,
\begin{equation}\label{eq:Hstart}
\begin{aligned}
H/\hbar & =H(\omega_1,\delta_1,a)/\hbar+H(\omega_2,\delta_2,b)/\hbar+\omega_r c^\dagger c\\
& +g_1(ac^\dagger+a^\dagger c)+g_2(bc^\dagger+b^\dagger c).
\end{aligned}
\end{equation}
In the dispersive limit (i.e. assuming $|\omega_r-\omega_\iota|\gg g_\iota,\delta_\iota$), this Hamiltonian is approximated by the effective Hamiltonian
\begin{equation}\label{eq:dispersive}
\begin{aligned}
H_d/\hbar = H'_1(|m\rangle,c)/\hbar + & H'_2(|n\rangle,c)/\hbar + \omega_r c^\dagger c \\
+ \sum_{j,k}\sqrt{(j+1)(k+1)} & J_{j,k} \left(|j,k+1\rangle\langle j+1,k|\right. \\
& +\left.|j+1,k\rangle\langle j,k+1|\right)
\end{aligned}
\end{equation}
where
\begin{equation}
H'_\iota(|k\rangle,c)/\hbar = \sum_k\tilde{\omega}_{\iota:k}|k\rangle\langle k| + \sum_k\chi_{\iota:k}c^\dagger c|k\rangle\langle k|
\end{equation}
and
\begin{align}
J_{j,k} & = \frac{g_1g_2(\omega_1+\omega_2+j\delta_1+k\delta_2-2\omega_r)}{2(\omega_1+j\delta_1-\omega_r)(\omega_2+k\delta_2-\omega_r)} \\
\chi_{\iota:k} & = \frac{g_\iota^2(\delta_\iota-\omega_\iota+\omega_r)}{(\omega_\iota+k\delta_\iota-\omega_r)(\omega_j+(k-1)\delta_\iota-\omega_r)} \\
\tilde{\omega}_{\iota:k} & = k\omega_\iota + \frac{\delta_\iota}{2}k(k-1)+\frac{kg_\iota^2}{\omega_\iota+(k-1)\delta_\iota-\omega_r}.\label{eq:lamb}
\end{align}
A further perturbative approximation replaces the coupling term with $J(a^\dagger b+ab^\dagger)$ where
\begin{equation}
J = \frac{g_1g_2(\omega_1+\omega_2-2\omega_r)}{2(\omega_1-\omega_r)(\omega_2-\omega_r)}.
\end{equation}

\section{Static coupling}\label{sec:alwaysZZ}

If $J\ll |\omega_1-\omega_2|$, we can make a further perturbation expansion to approximate the diagonal terms of the Hamiltonian in the qubit subspace as
\begin{equation}
H_{\mathrm{eff}}/\hbar=\Xi_m^{(1)}Z_1/4+\Xi_m^{(2)}Z_2/4+\zeta_mZ_1Z_2/4
\end{equation}
where $\Xi_m^{(1)}$, $\Xi_m^{(2)}$, and $\zeta_m$ are real functions of photon number $m$, and $Z_\iota$ is a $Z$ Pauli matrix acting on qubit $\iota$, i.e. $Z_1=Z\otimes I$ and $Z_2=I\otimes Z$. The energies of the uncoupled eigenstates $|jkm\rangle$ are $E_{jkm}/\hbar = (\chi_{1:j}+\chi_{2:k})m + \tilde{\omega}_{1:j} + \tilde{\omega}_{2:k}$, and the second order corrections are
\begin{align}
E_{01m}^{(2)}/\hbar & = \frac{J^2}{E_{01m}/\hbar-E_{10m}/\hbar} \\
E_{10m}^{(2)}/\hbar & = -E_{01m}^{(2)}/\hbar \\
E_{11m}^{(2)}/\hbar & = \frac{2J^2}{E_{11m}/\hbar-E_{02m}/\hbar} + \frac{2J^2}{E_{11m}/\hbar-E_{20m}/\hbar}.
\end{align}
The effective Hamiltonian in the qubit and $m$ photon subspace is $H_\mathrm{eff}^{(m)}=\mathrm{diag}(E_{00m},E_{01m}+E_{01m}^{(2)},E_{10m}+E_{10m}^{(2)},E_{11m}+E_{11m}^{(2)})$ from which we find $\Xi_m^{(\iota)}=\mathrm{Tr}(H_\mathrm{eff}^{(m)}Z_\iota)/\hbar$ and $\zeta_m=\mathrm{Tr}(H_\mathrm{eff}^{(m)}Z_1Z_2)/\hbar$.
After simplification, and neglecting the Lamb shift (Eq.~\ref{eq:lamb}), 
\begin{equation}
\zeta_0 \approx -\frac{2J^2(\delta_1+\delta_2)}{(\delta_1+\omega_1-\omega_2)(\delta_2-\omega_1+\omega_2)}.
\end{equation}

We observe that the coefficients $\Xi_m^{(\iota)}$ are well approximated by their values at $m=0$, since the fractional error $(\Xi_m^{(\iota)}-\Xi_0^{(\iota)})/\Xi_0^{(\iota)}$ is a few percent or less for $m<50$ and typical parameters (see App.~\ref{sec:param}). However, the fractional error for $\zeta_m$ approaches $40\%$ at $m=50$. This is another manifestation of the nonlinearity that enters at fourth order in the qubit-resonator coupling. Nevertheless, at low photon numbers $n\approx 5$, the fractional error is a few percent and the approximation appears to be acceptable.

\section{Master equation}\label{sec:solution}

We begin with the Hamiltonian in the dispersive limit (Eq.~\ref{eq:dispersive}) and retain only the qubit subspace of the non-linear oscillators as given in Sec.~\ref{sec:alwaysZZ},
\begin{equation}
\begin{aligned}
H''/\hbar & = \omega_r c^\dagger c + \Xi_0^{(1)}Z_1/4 + \Xi_0^{(2)}Z_2/4 \\
& + \sum_{j,k} \chi_{jk}c^\dagger c|jk\rangle\langle jk| + \zeta_0Z_1Z_2/4
\end{aligned}
\end{equation}
where $\chi_{jk}=\chi_{1:j}+\chi_{2:k}$. Discarding higher levels is acceptable because we drive the system far from resonance, so additional levels of each non-linear oscillator undergo a diagonal unitary evolution in the dressed basis. We add a term $\hbar\epsilon(t)(c+c^\dagger)$
driving the bus resonator where $\epsilon(t)=\epsilon_I(t)\cos(\omega_d t)+\epsilon_Q(t)\sin(\omega_d t)$. Applying a frame transformation 
\begin{equation}
R(t)=e^{-it\left(-\Xi_0^{(1)}a^\dagger a/2-\Xi_0^{(2)}b^\dagger b/2+\omega_dc^\dagger c\right)}
\end{equation}
and making the rotating wave approximation $\mathrm{exp}(\pm i2\omega_d)\mapsto 0$,
\begin{equation}\label{eq:closedformH}
\begin{aligned}
H^R/\hbar & = (\omega_r-\omega_d) c^\dagger c + \zeta_0Z_1Z_2/4  + \sum_{j,k} \chi_{jk}c^\dagger c|jk\rangle\langle jk| \\
& + \frac{\epsilon_I(t)}{2}(c+c^\dagger)+\frac{\epsilon_Q(t)}{2}(-ic+ic^\dagger).
\end{aligned}
\end{equation}
Our goal is to solve the master equation
\begin{equation}\label{eq:master}
\dot{\rho} = -\frac{i}{\hbar}[H^R,\rho]+\kappa D[c]\rho
\end{equation}
where $\kappa$ is the resonator loss rate and
\begin{equation}\label{eq:dissipator}
D[c]\rho := (2c\rho c^\dagger-c^\dagger c\rho-\rho c^\dagger c)/2.
\end{equation}

Write the density operator for the coupled system as
\begin{equation}\label{eq:coupledform}
\rho = \sum_{jklm} C_{jk,lm}\otimes |jk\rangle\langle lm|
\end{equation}
where $j$, $k$, $l$, $m$ label states of the total nonlinear oscillator Hilbert space and $C_{jk,lm}$ are blocks of the density operator but are not density operators themselves. Since $\rho$ is a valid density operator, the blocks satisfy $C_{jk,lm}=C_{lm,jk}^\dagger$ and $\sum_{jk}\mathrm{Tr}\ C_{jk,jk}=1$. The partial trace over the bus resonator gives the reduced density operator of the non-linear oscillator Hilbert space whose matrix elements are $(\mathrm{Tr}_{\mathrm{cav}}\rho)_{jk,lm}=\mathrm{Tr}\ C_{jk,lm}$. Each $C_{jk,lm}$ encodes the initial state of the qubit-bus system.

Substituting Eq.~\ref{eq:coupledform} into Eq.~\ref{eq:master}, the diagonal blocks satisfy
\begin{widetext}
\begin{equation}
\dot{C}_{jk,jk} = -i(\omega_r-\omega_d+\chi_{jk})[c^\dagger c,C_{jk,jk}]-i\frac{\epsilon_I(t)}{2}[c+c^\dagger,C_{jk,jk}]-i\frac{\epsilon_Q(t)}{2}[-ic+ic^\dagger,C_{jk,jk}]+\kappa D[c]C_{jk,jk}
\end{equation}
and the remaining blocks satisfy
\begin{equation}
\begin{aligned}
\dot{C}_{jk,lm} & = -i(\omega_r-\omega_d)[c^\dagger c,C_{jk,lm}]-i\zeta_0((-1)^{j+k}-(-1)^{l+m})/4 - i\chi_{jk}c^\dagger cC_{jk,lm}+i\chi_{lm}C_{jk,lm}c^\dagger c \\
&-i\frac{\epsilon_I(t)}{2}[c+c^\dagger,C_{jk,lm}]-i\frac{\epsilon_Q(t)}{2}[-ic+ic^\dagger,C_{jk,lm}]+\kappa D[c]C_{jk,lm}.
\end{aligned}
\end{equation}
\end{widetext}
We make use of a generalized P-representation \cite{drummond80}
\begin{equation}
C_{jk,lm} = \int d\alpha^2 \int d\beta^2 \Lambda(\alpha,\beta) P_{jk,lm}(\alpha,\beta)
\end{equation}
where $\Lambda(\alpha,\beta)=|\alpha\rangle\langle\beta^\ast|/\langle\beta^\ast|\alpha\rangle$, which gives rise to the operator correspondences
\begin{align}
cC_{jk,lm} & \leftrightarrow \alpha P_{jk,lm}(\alpha,\beta) \\
c^\dagger C_{jk,lm} & \leftrightarrow (\beta-\partial_\alpha) P_{jk,lm}(\alpha,\beta) \\
C_{jk,lm}c^\dagger & \leftrightarrow \beta P_{jk,lm}(\alpha,\beta) \\
C_{jk,lm}c & \leftrightarrow (\alpha-\partial_\beta) P_{jk,lm}(\alpha,\beta).
\end{align}
The generalized P-representation is required to vanish rapidly enough that the integral exists as the magnitude of $\alpha$ and $\beta$ approach infinity.
The corresponding equations
\begin{align}
\dot{P}_{jk,jk} & = \partial_\alpha\left( \tilde{\Delta}_{jk}\alpha + \frac{i}{2}\tilde{\epsilon}(t)\right)P_{jk,jk} \\
& + \partial_\beta\left( \tilde{\Delta}_{jk}^\ast\beta - \frac{i}{2}\tilde{\epsilon}^\ast(t)\right)P_{jk,jk} \nonumber \\
\dot{P}_{jk,lm} & = \partial_\alpha\left( \tilde{\Delta}_{jk}\alpha + \frac{i}{2}\tilde{\epsilon}(t)\right)P_{jk,lm} \\
& + \partial_\beta\left( \tilde{\Delta}_{lm}^\ast\beta - \frac{i}{2}\tilde{\epsilon}^\ast(t)\right)P_{jk,lm} \nonumber \\
& -i(\chi_{jk}-\chi_{lm})\alpha\beta P_{jk,lm} \nonumber \\
& -i\zeta_0((-1)^{j+k}-(-1)^{l+m})/4 P_{jk,lm} \nonumber
\end{align}
describe the evolution of the generalized density $P_{jk,lm}$ where we have defined
$\tilde{\Delta}_{jk}=i(\omega_r-\omega_d+\chi_{jk})+\kappa/2$ and
$\tilde{\epsilon}(t) = \epsilon_I(t)+i\epsilon_Q(t)$.

Let $f_{jk,lm}(\alpha,\beta)=\delta^2(\alpha-\alpha_{jk}(t))\delta^2(\beta-\alpha_{lm}^\ast(t))$. Substituting the ansatz,
\begin{align}
P_{jk,jk}(\alpha,\beta) & = p_{jk}f_{jk,jk}(\alpha,\beta), \\
P_{jk,lm}(\alpha,\beta) & =
e^{i\mu_{jk,lm}(t)}f_{jk,lm}(\alpha,\beta),
\end{align}
gives a deterministic evolution governed by
\begin{align}
\dot{\alpha}_{jk} & = -\tilde{\Delta}_{jk}\alpha_{jk} - \frac{i}{2}\tilde{\epsilon}(t) \\ 
\dot{\mu}_{jk,lm} & = (\chi_{lm}-\chi_{jk})\alpha_{lm}^\ast\alpha_{jk} \\
& + \zeta_0((-1)^{l+m}-(-1)^{j+k})/4.\nonumber
\end{align}
These equations have the solution
\begin{align}
\alpha_{jk}(t)& =\alpha_{jk}(0)e^{-\tilde{\Delta}_{jk}t} \\
& -\frac{i}{2}\int_0^t e^{-\tilde{\Delta}_{jk}(t-t')}\tilde{\epsilon}(t')dt' \nonumber\\
\mu_{jk,lm}(t) & = \mu_{jk,lm}(0) \nonumber \\
& + (\chi_{lm}-\chi_{jk})\int_0^t\alpha_{lm}^\ast(t')\alpha_{jk}(t')dt' \\
& + \zeta_0((-1)^{l+m}-(-1)^{j+k})t/4.\nonumber
\end{align}
This is equivalent to stating that
\begin{equation}
\begin{aligned}
& \rho(t) = \sum_{jk}p_{jk}|\alpha_{jk}(t),jk\rangle\langle\alpha_{jk},jk| \\
& + \sum_{jk\neq lm} \frac{e^{i\mu_{jk,lm}(t)}}{\langle\alpha_{lm}(t)|\alpha_{jk}(t)\rangle}|\alpha_{jk}(t),jk\rangle\langle\alpha_{lm}(t),lm|
\end{aligned}
\end{equation}
from which the reduced qubit and bus resonator states can be found.

\section{Geometric phase}\label{sec:geometric}

It is not immediately obvious that the integral in Eq.~\ref{eq:phaseevol} can be expressed as a sum of dynamical and geometric phases. Begin by expanding the resonator response as in Eq.~\ref{eq:adiabatic} to order $M=2$. Suppose that $\tilde{\epsilon}(0)=\tilde{\epsilon}^{(1)}(0)=0$ and that $\tilde{\epsilon}(t)$ varies sufficiently slowly that the last term of Eq.~\ref{eq:adiabatic} can be neglected. Substituting the approximate responses into Eq.~\ref{eq:phaseevol}, we find that
\begin{equation}
\begin{aligned}
\mathrm{Re}\left[(\bar{\chi}_{jk}-\bar{\chi}_{lm})\int_0^t\right. & \left.\left[\alpha_{lm}^{(0)}(t')\right]^\ast \alpha_{jk}^{(0)}(t')dt'\right] \\
& \approx \gamma_d(t) + \gamma_g(t) + \gamma_r(t)
\end{aligned}
\end{equation}
where the dynamical phase is given by
\begin{equation}
\gamma_d(t) = \mathrm{Re}\left[\frac{\bar{\chi}_{jk}-\bar{\chi}_{lm}}{4\tilde{\Delta}_{lm}^\ast\tilde{\Delta}_{jk}}\int_0^t |\tilde{\epsilon}(\tau)|^2d\tau\right],\label{eq:dynamicphase}
\end{equation}
the geometric phase is given by
\begin{equation}
\gamma_g(t) = -\mathrm{Re}\int_0^t \frac{\tilde{\Delta}_{jk}\tilde{\epsilon}(\tau)\dot{\tilde{\epsilon}}^{\ast}(\tau)+\tilde{\Delta}_{lm}^\ast\tilde{\epsilon}^\ast(\tau)\dot{\tilde{\epsilon}}(\tau)}{4(\tilde{\Delta}_{lm}^\ast)^2(\tilde{\Delta}_{jk})^2}d\tau,\label{eq:geometricphase}
\end{equation}
and the residual phase term
\begin{equation}
\gamma_r(t) = \mathrm{Re}\left[(\tilde{\chi}_{jk}-\tilde{\chi}_{lm})\int_0^t \frac{|\dot{\tilde{\epsilon}}(\tau)|^2}{4(\tilde{\Delta}_{lm}^\ast)^2(\tilde{\Delta}_{jk})^2}d\tau\right]\label{eq:residualphase}
\end{equation}
is assumed to be negligible in the adiabatic regime. In the limit of $\kappa\mapsto 0$, we can use the identities $\mathrm{Re}\left[-ic\right]=\mathrm{Re}\left[ic^\ast\right]=\mathrm{Im}\left[c\right]$ and $\oint_\lambda\mathrm{Im}\left[\tilde{\epsilon}(\lambda)^\ast \frac{d\tilde{\epsilon}(\lambda)}{d\lambda} d\lambda\right]=2A_{\tilde{\epsilon}}$ to confirm that
\begin{equation}
\gamma_g(t) = \frac{(\bar{\chi}_{jk}-\bar{\chi}_{lm})(2\Delta+\bar{\chi}_{jk}+\bar{\chi}_{lm})}{2(\Delta+\bar{\chi}_{jk})^2(\Delta+\bar{\chi}_{lm})^2}A_{\tilde{\epsilon}}
\end{equation}
where $A_{\tilde{\epsilon}}$ is the (signed) area enclosed by $\tilde{\epsilon}(t)$.

\section{Parameter values}\label{sec:param}

The resonator shifts for each qubit state are given by
\begin{align}
\chi_{00} & = -\frac{g_1^2}{\omega_1-\omega_r} - \frac{g_2^2}{\omega_2-\omega_r} \\
\chi_{01} & = -\frac{g_1^2}{\omega_1-\omega_r} + \frac{g_2^2}{\omega_2-\omega_r} - \frac{2g_2^2}{\omega_2-\omega_r+\delta_2} \\
\chi_{10} & = -\frac{g_2^2}{\omega_2-\omega_r} + \frac{g_1^2}{\omega_1-\omega_r} - \frac{2g_1^2}{\omega_1-\omega_r+\delta_1} \\
\chi_{11} & = \frac{g_1^2}{\omega_1-\omega_r} - \frac{2g_1^2}{\omega_1-\omega_r+\delta_1} \\
& + \frac{g_2^2}{\omega_2-\omega_r} - \frac{2g_2^2}{\omega_2-\omega_r+\delta_2}
\end{align}
at second order in the bus coupling strength. We choose experimentally reasonable parameter values $g_\iota/2\pi=120\ \mathrm{MHz}$, $\delta_\iota/2\pi=-300\ \mathrm{MHz}$, and $\omega_r/2\pi=7\ \mathrm{GHz}$, and consider two different sets of qubit frequencies that we call ``low'' and ``high''. The words are merely labels; neither set of parameters is physically extreme.

For ``low'' qubit frequencies, let $\omega_1/2\pi=5\ \mathrm{GHz}$ and  $\omega_2/2\pi=4.95\ \mathrm{GHz}$. The corresponding resonator shifts are approximately $\chi_{00}/2\pi\approx 14.22\ \mathrm{MHz}$, $\chi_{01}/2\pi\approx 12.43\ \mathrm{MHz}$, $\chi_{10}/2\pi\approx 12.34\ \mathrm{MHz}$, and $\chi_{11}/2\pi\approx 10.55\ \mathrm{MHz}$. From these shifts we compute the shift of the dressed resonator frequency to be $\bar{\chi}_{01}/2\pi\approx 1.79\ \mathrm{MHz}$, $\bar{\chi}_{10}/2\pi\approx 1.88\ \mathrm{MHz}$, and $\bar{\chi}_{11}/2\pi\approx 3.67\ \mathrm{MHz}$
as a function of qubit state $|jk\rangle$. The effective coupling strength is $J/2\pi\approx -7.11\ \mathrm{MHz}$ which leads to a value of $\zeta_0/2\pi\approx 694\ \mathrm{kHz}$. Therefore, the free evolution can maximally entangle in roughly $\pi/\zeta_0\approx 720 \mu\mathrm{s}$.

For ``high'' qubit frequencies, $\omega_1/2\pi=5.6\ \mathrm{GHz}$ and  $\omega_2/2\pi=5.55\ \mathrm{GHz}$. The resonator shifts are $\chi_{00}/2\pi\approx 20.22\ \mathrm{MHz}$, $\chi_{01}/2\pi\approx 16.81\ \mathrm{MHz}$, $\chi_{10}/2\pi\approx 16.59\ \mathrm{MHz}$, and $\chi_{11}/2\pi\approx 13.18\ \mathrm{MHz}$. The dressed resonator frequency shifts are $\bar{\chi}_{01}/2\pi\approx 3.40\ \mathrm{MHz}$, $\bar{\chi}_{10}/2\pi\approx 3.63\ \mathrm{MHz}$, and $\bar{\chi}_{11}/2\pi\approx 7.04\ \mathrm{MHz}$. The effective coupling strength is $J/2\pi\approx -10.11\ \mathrm{MHz}$ which leads to a value of $\zeta_0/2\pi\approx 1.40\ \mathrm{MHz}$. Free evolution can maximally entangle in roughly $\pi/\zeta_0\approx 357 \mu\mathrm{s}$.

Resonators have quality factors of about $Q\approx 150,000$ and dressed frequencies of $(\omega_r+\chi_{00})/2\pi\approx 7$ GHz, which corresponds to a loss rate of $\kappa/2\pi\approx 50$ kHz and $2\pi/\kappa\approx 20\ \mu\mathrm{s}$. The measured shift of the resonator $\bar{\chi}_{01}/2\pi\approx\bar{\chi}_{10}/2\pi$ is on the order of MHz. The drive detuning $\Delta/2\pi$ can be varied over tens of MHz with peak Rabi rates up to the order of hundreds of MHz.

\section{Average gate fidelity}\label{sec:fidelity}

The average gate fidelity is one way to quantify how well a quantum operation approximates a given gate \cite{NC}. The average fidelity between a quantum
operation ${\cal E}(\rho)$ and a unitary gate $U$ is given by
\begin{equation}
{\cal F}_{\mathrm{A}}(U,{\cal E})=\frac{d{\cal F}_{\mathrm{E}}(U,{\cal E})+1}{d+1}
\end{equation}
where $d$ is the dimension of the Hilbert space and ${\cal F}_{\mathrm{E}}(U,{\cal E})$ is the entanglement fidelity.
The entanglement fidelity is given by
\begin{equation}
{\cal F}_{\mathrm{E}}(U,{\cal E}) = \frac{1}{d^3}\sum_{P\in G_d} \mathrm{Tr}[PU^\dagger{\cal E}(P)U]
\end{equation}
where $G_d$ is the group of $d\times d$ Pauli matrices modulo phase \cite{nielsen02}.

A controlled-Z gate is locally equivalent to the gate $U:=\mathrm{exp}(-i\pi/4 Z\otimes Z)$; specifically, $\Lambda(Z)=e^{i\pi/4}(S^\dagger\otimes S^\dagger)U$ where $S:=\mathrm{diag}(1,i)$. Let ${\cal E}(\rho)$ be the quantum operation given implicitly by the solution in App.~\ref{sec:solution} and let ${\cal F}(\rho):=(X\otimes X)\rho(X\otimes X)$. The average fidelity between $U$ and the composite sequence ${\cal G}:={\cal F}\circ {\cal E}\circ {\cal F}\circ {\cal E}$ is given by
\begin{align}
{\cal F}_{\mathrm{A}}(U,{\cal G}) & = \frac{2}{5} + \frac{e^{-2\mathrm{Im}[\mu_{00,11}]}+e^{-2\mathrm{Im}[\mu_{01,10}]}}{10} \\
& - \frac{1}{5}e^{-\mathrm{Im}[\mu_{00,10}-\mu_{01,11}]}\sin(\mathrm{Re}[\mu_{00,10}-\mu_{01,11}]) \nonumber \\
& - \frac{1}{5}e^{-\mathrm{Im}[\mu_{00,01}-\mu_{10,11}]}\sin(\mathrm{Re}[\mu_{00,01}-\mu_{10,11}]).\nonumber
\end{align}
The average gate infidelity is $1-{\cal F}_{\mathrm{A}}(U,{\cal G})$.

\section{Numerical solution}\label{sec:numerics}

Here we describe numerical solution of the master equation. We begin with Eq.~\ref{eq:Hstart} and add a drive term $\hbar\epsilon(t)(c+c^\dagger)$. Apply the frame transformation
$R'(t)=\mathrm{exp}(-i\omega_dt(a^\dagger a + b^\dagger b + c^\dagger c))$
and make the rotating wave approximation on the drive terms
to find
\begin{equation}\label{eq:numericH}
\begin{aligned}
H/\hbar & =H(\omega_1-\omega_d,\delta_1,a)+H(\omega_2-\omega_d,\delta_2,b)\\
& +(\omega_r-\omega_d)c^\dagger c\\
& +g_1(ac^\dagger+a^\dagger c)+g_2(bc^\dagger+b^\dagger c) \\
& +\frac{\epsilon_I(t)}{2}(c+c^\dagger)+\frac{\epsilon_Q(t)}{2}(-ic+ic^\dagger).
\end{aligned}
\end{equation}
Using adaptive 8(9) Runga-Kutta-Dormand-Prince numerical integration, we solve the master equation
\begin{equation}
\dot{\rho} = -\frac{i}{\hbar}[H,\rho] + \kappa D[c]\rho
\end{equation}
where $D[c]\rho$ is given by Eq.~\ref{eq:dissipator}. Differences in the local frame between Eq.~\ref{eq:numericH} and Eq.~\ref{eq:closedformH} cannot produce entanglement or change the resonator response. The numerical solutions have been used to verify the closed-form solutions for relatively low numbers of resonator photons $\langle n\rangle\approx 5-10$ in the case of constant and spline drive shapes, and we have found that the solutions are consistent. For the case of spline drive shapes, the solutions are almost identical.

\end{document}